\documentclass{ws-procs975x65}

\begin{document}

%to switch ON running title
%\markboth{M. Daszkiewicz, J. Lukierski and M. Woronowicz}{Noncommutative translations and $\star$-product formalism}
%\wstoc{Noncommutative translations and $\star$-product formalism}{M. Daszkiewicz, J. Lukierski and M. Woronowicz}

\title{NONCOMMUTATIVE TRANSLATIONS AND $\star$-PRODUCT FORMALISM\footnote{Supported by KBN grant 1P03B01828.}
\footnote{Presented by J. Lukierski, e-mail:\ lukier@ift.uni.wroc.pl}}

\author{Marcin Daszkiewicz, Jerzy Lukierski and Mariusz Woronowicz}

\address{Institute of Theoretical Physics\\ Wroc{\l}aw University  pl. Maxa
Borna 9, 50-206 Wroc{\l}aw, Poland}

\begin{abstract}

    We consider the noncommutative space-times with Lie-algebraic
    noncommutativity (e.g. $\kappa$-deformed Minkowski space).
    In the framework with classical fields  we extend the $\star$-product
   in order to   represent the noncommutative
    translations in terms of commutative ones. We show the translational
    invariance of noncommutative bilinear action with local product of
    noncommutative fields. The quadratic noncommutativity is also
    briefly discussed.

\end{abstract}

\bodymatter

\vspace{0.5cm}

%%%%\section{xx}
\noindent

In noncommutative space-time, in general case, the translations are
also noncommutative. The aim of this note is to study the
translational invariance of local noncommutative actions.

The noncommutative Minkowski space
\begin{equation}
\label{nc1} [\,{\hat x}_{\mu},{\hat x}_{\nu}\,] = \frac{i}{\kappa^2}
\theta(\kappa {\hat x})\;,
\end{equation}
where we choose $(\widetilde{x} = \kappa{\hat x})$
\begin{equation}
\label{nc2} \theta (\widetilde{x}) = \theta_{\mu \nu}^{(1)
\lambda}{\widetilde{x}}_{\lambda} + \theta_{\mu \nu}^{(2) \lambda
\rho}{\widetilde{x}}_{\lambda} {\widetilde{x}}_{\rho}\;,
\end{equation}
is invariant under the translations
\begin{equation}
\label{nc3} {\hat x}_{\mu} \rightarrow {{\hat x}'}_{\mu} = {\hat
x}_{\mu} + {\hat v}_{\mu}\;,
\end{equation}
if
\begin{equation}
\label{nc4} [\,{\hat v}_{\mu},{\hat v}_{\nu}\,] =
\frac{i}{\kappa}\theta_{\mu \nu}^{(1) \lambda}{\hat v}_{\lambda} +
i\theta_{\mu \nu}^{(2) \lambda \rho}{\hat v}_{\lambda} {\hat
v}_{\rho} \;,
\end{equation}
\begin{equation}
[\,{\hat x}_{\mu},{\hat v}_{\nu}\,] = \frac{i}{2}\theta_{\mu
\nu}^{(2) \lambda \rho} \left ({\hat x}_{\lambda} {\hat v}_{\rho} +
{\hat x}_{\rho} {\hat v}_{\lambda}\right )\;. \label{nc5}
\end{equation}
If the relation (\ref{nc3}) describes a coproduct from the relation
(\ref{nc5}) follows that for quadratic deformations  such a
coproduct is a braided one (see also \cite{majid,chryss}). Contrary
to the recent proposal\cite{arzano} , in Lie-algebraic case the
formula (\ref{nc3}) implies that the noncommutative translations are
represented by standard Hopf-algebraic coproduct. It should be
recalled that such standard coproduct describes the translation
sector of quantum $\kappa$-Poincar\'{e} group \cite{kappa} .

Let us choose firstly in (\ref{nc1}-\ref{nc5}) $\theta_{\mu
\nu}^{(1) \lambda} \ne 0$ and $\theta_{\mu \nu}^{(2) \rho \lambda} =
0$ (Lie-algebraic case). In such a case the relations (\ref{nc1})
and (\ref{nc4}-\ref{nc5}) describe two commuting copies of Lie
algebra with the structure constant $\theta_{\mu \nu}^{(1)
\lambda}$. \\
Using CBH formula for the multiplication of the group
elements of the corresponding Lie group
 (see e.g. \cite{kath})
\begin{equation}
\label{nc6}
{\rm e}^{i\alpha^{\mu}{\hat x}_{\mu}}{\rm
e}^{i\beta^{\mu}{\hat x}_{\mu}} = {\rm
e}^{i\gamma^{\mu}(\alpha,\beta){\hat x}_{\mu}}\;,
\end{equation}
where
\begin{equation}
\label{nc7}
\gamma^{\mu}(\alpha,\beta) = \alpha^{\mu} + \beta^{\mu} +
\frac{1}{\kappa}\theta_{\nu \rho}^{(1) \mu}\alpha^{\nu}\beta^{\rho}
+ \frac{1}{12\kappa^2}\theta_{\rho \tau}^{(1) \mu}\theta_{\lambda
\nu}^{(1) \rho}(\alpha^{\tau}\alpha^{\lambda}\beta^{\nu} +
\beta^{\tau}\beta^{\lambda}a^{\nu}) + \cdots\;,
\end{equation}
one can introduce the following $\star$-product of the classical
exponentials
\begin{equation}
\label{nc8} {\rm e}^{i\alpha^{\mu}{x}_{\mu}}\star {\rm
e}^{i\beta^{\mu}{x}_{\mu}} = {\rm
e}^{i\gamma^{\mu}(\alpha,\beta){x}_{\mu}}\;.
\end{equation}
For two arbitrary classical field the formula (\ref{nc8}) generates
the following $\star$-multiplication
\begin{eqnarray}
\label{nc9} \phi(x)\star \chi(x) &=& \lim_{y,z\rightarrow x}\phi(y)~
{\rm exp}
\Big(
%\left(
 ix_{\mu}\widetilde{\gamma}^{\mu}
%\left
\Big(
{\overleftarrow{\frac{\partial}{\partial y}}},
\overrightarrow{\frac{\partial}{\partial z}}
%\right)
%\right)
\Big)
\Big)
 \chi(z)
\cr &=& \int d^4x_1d^4x_2K(x;x_1,x_2)\phi(x_1)\chi(x_2)\;,
\end{eqnarray}
where $\widetilde{\gamma}^{\mu}(\alpha,\beta) =
{\gamma}^{\mu}(\alpha,\beta) - \alpha^{\mu} - \beta^{\mu}$ and the
nonlocal kernel $K(x;y,z)$ describes the bidifferential operator of
infinite order. \\
The product of two noncommutative fields ${\hat \phi}({\hat x},{\hat
v}){\hat \chi}({\hat x},{\hat v})$ is represented as the product of
two commuting $\star$-products (\ref{nc9})
\begin{eqnarray}
\label{nc10} {\phi}({x},{v})\star {\chi}({x},{v}) &=&
\lim_{y,z\rightarrow x}~\lim_{u,w\rightarrow v}  \phi(y,u)~{\rm exp}
%\left(
\Big(
 ix_{\mu}\widetilde{\gamma}^{\mu}
%\left(
\Big(
{\overleftarrow{\frac{\partial}{\partial y}}},
\overrightarrow{\frac{\partial}{\partial z}}
%\right)
\Big)
+
%\right.
\cr
&+&iv_{\mu}
%\left.
 \widetilde{\gamma}^{\mu}
%\left(
\Big(
{\overleftarrow{\frac{\partial}{\partial y}}},
\overrightarrow{\frac{\partial}{\partial z}}
%\right)
\Big)
%\right)
\Big)
\chi(z,w)\;.
\end{eqnarray}
For ${\hat \phi}({\hat x},{\hat v}) = {\hat \phi}({\hat x} + {\hat
v})$ and ${\hat \chi}({\hat x},{\hat v}) = {\hat \chi}({\hat x} +
{\hat v})$ one can put on r.h.s. of (\ref{nc10}) $\phi(y,u) =
\phi(y+u)$ and $\chi(z,w) = \chi(z+w)$. Using
$\frac{\partial}{\partial y} = \frac{\partial}{\partial u}$,
$\frac{\partial}{\partial z} = \frac{\partial}{\partial w}$ and
(\ref{nc9}), one gets
\begin{eqnarray}
\label{nc11} {\phi}({x}+{v})\star {\chi}({x}+{v}) &=&
\lim_{y,z\rightarrow x}\phi(y+v)~ {\rm exp}\left (
i(x_{\mu}+v_{\mu})\widetilde{\gamma}^{\mu} \left
({\overleftarrow{\frac{\partial}{\partial y}}},
\overrightarrow{\frac{\partial}{\partial z}}\right)\right )\chi(z+v)
\cr &=& \int d^4x_1d^4x_2K(x+v;x_1,x_2)\phi(x_1)\chi(x_2)\;.
\end{eqnarray}
We introduce the noncommutative integration satisfying the
relation
\begin{equation}
\label{nc12}
\int d^{4}{\hat x}{\hat F}({\hat x}) = \int
d^{4}x \, \mu(x)\, {F}({x})\;,
\end{equation}
where $\mu(x)$ is adjusted by the cyclic property of the
noncommutative integral when ${\hat{F}}(\hat{x}) =
{\hat{\phi}}(\hat{x}) {\hat{\chi}}(\hat{x})$ (see e.g.
\cite{dijomo}). The translational invariance of standard integration
and the formula (\ref{nc11}) implies that
\begin{equation}
\label{nc13}
\int d^{4}{\hat x}{\hat \phi}({\hat x} + {\hat v}){\hat
\chi}({\hat x}+ {\hat v}) = \int d^{4}{\hat x}{\hat \phi}({\hat
x}){\hat \chi}({\hat x})\;.
\end{equation}
The formula (\ref{nc13}) describes explicitly the translational
invariance of bilinear action under noncommutative coordinate shifts
(\ref{nc3}).

The star product (\ref{nc8}) describes the multiplication of
nonordered noncomutative plane waves. In particular case of
Lie-algebraic deformation, it is useful to consider the
noncommutative plane waves ordered in particular way. For example,
if we assume that the commutator (\ref{nc1}) describes
$\kappa$-deformed Minkowski space \cite{kappa,majid2}
\begin{equation}
\label{nc14} [\,{\hat x}_{0},{\hat x}_{i}\,] = \frac{i}{\kappa}{\hat
x}_{i}\;\;,\;\;[\,{\hat x}_{i},{\hat x}_{j}\,] = 0\;,
\end{equation}
one can introduce the normally ordered exponentials
\cite{majid2,kosinski} 
\begin{equation}
\label{nc15} :{\rm e}^{ip^{\mu}{\hat x}_{\mu}}: = {\rm
e}^{ip^{0}{\hat x}_{0}} {\rm e}^{ip^{i}{\hat x}_{i}}\;.
\end{equation}
Using the relation
\begin{equation}
\label{nc16} {\rm e}^{ip^{\mu}{\hat x}_{\mu}} = {\rm
e}^{i\widetilde{p}^{0}{\hat x}_{0}} {\rm e}^{i\widetilde{p}^{i}{\hat
x}_{i}}\;,
\end{equation}
where
\begin{equation}
\label{nc17} \widetilde{p}^0 = p^0\;\;,\;\;\widetilde{p}^i =
\frac{\kappa}{p_0}\left (1-{\rm e}^{-\frac{p^0}{\kappa}} \right
)p^i\;,
\end{equation}
one can translate the CBH star product (\ref{nc8}) into the standard
star product, used in $\kappa$-deformed field theory
\cite{kosinski,9} , which is homomorphic to the
multiplication of normally ordered exponentials. In fact, there is
an infinite number of ways to define the star product, homomorphic
to noncommutative multiplication rule, which is related by various
nonlinear transformations of the four-momentum variable (see
e.g.\cite{kosinski,10}).

Finally, let us consider quadratic deformations of Minkowski space.
If we choose in (\ref{nc1}-\ref{nc5}) $\theta_{\mu \nu}^{(1)
\lambda} = 0$ and $\theta_{\mu \nu}^{(2) \rho \lambda} \ne 0$, the
star product representing the noncommutative translations has to
take into consideration the braiding (relation (\ref{nc5})), i.e.
contrary to the formula (\ref{nc10}), it does not factorize into the
product of two identical $\star$-products. If we correctly, however,
introduce one "big" star product representing the noncommutativity
given by (\ref{nc1}), (\ref{nc4}) and (\ref{nc5}), it is possible to
represent the noncommutative quadratic translations by the classical
ones. In order to show the translational invariance of corresponding
noncommutative local field theory, one has to find for quadratic
deformations the counterpart of the relation (\ref{nc12}), which is
less obvious than in the case of Lie-algebraic space-time
commutation relations.

\end{document}